\documentclass[journal=jacsat,manuscript=article]{achemso}

\usepackage[version=3]{mhchem} 

\usepackage{natbib}
\usepackage{graphicx}
\usepackage[utf8]{inputenc}
\usepackage{mathptmx}
\usepackage{epsfig,amsopn}
\usepackage{graphicx}
\usepackage{amsmath,amssymb}
\usepackage{natbib,color}
\usepackage{braket}
\usepackage{float}
\usepackage{enumerate}
\usepackage[normalem]{ulem}
\usepackage{todonotes}
\allowdisplaybreaks[1]

\newcommand{\eref}[1]{Eq.~(\ref{#1})}%
\newcommand{\fref}[1]{Fig.~\ref{#1}} %

\def\bea{\begin{eqnarray}}
\def\eea{\end{eqnarray}}

\author{Ofir Tal-Friedman}
\affiliation
{School of Physics \& Astronomy, Raymond and Beverly Sackler Faculty of Exact Sciences, Tel Aviv University, Tel Aviv 6997801, Israel}
\author{Arnab Pal}
\affiliation[Unknown University]
{School of Chemistry, The Center for Physics and Chemistry of Living Systems, \& The Mark Ratner Institute for Single Molecule Chemistry, Tel Aviv University, Tel Aviv 6997801, Israel}
\email{arnabpal@mail.tau.ac.il}
\author{Amandeep Sekhon}
\affiliation[Unknown University]
{School of Chemistry, The Center for Physics and Chemistry of Living Systems, \& The Mark Ratner Institute for Single Molecule Chemistry, Tel Aviv University, Tel Aviv 6997801, Israel}
\author{Shlomi Reuveni}
\affiliation[Unknown University]
{School of Chemistry, The Center for Physics and Chemistry of Living Systems, \& The Mark Ratner Institute for Single Molecule Chemistry, Tel Aviv University, Tel Aviv 6997801, Israel}
\email{shlomire@tauex.tau.ac.il}
\author{Yael Roichman}
\affiliation
{School of Physics \& Astronomy, Raymond and Beverly Sackler Faculty of Exact Sciences, Tel Aviv University, Tel Aviv 6997801, Israel}
\alsoaffiliation[Unknown University]
{School of Chemistry, The Center for Physics and Chemistry of Living Systems, \& The Mark Ratner Institute for Single Molecule Chemistry, Tel Aviv University, Tel Aviv 6997801, Israel}
\email{roichman@tauex.tau.ac.il}

\title[An \textsf{achemso} demo]
  {Experimental  realization of diffusion with stochastic resetting}

\abbreviations{IR,NMR,UV}
\keywords{American Chemical Society, \LaTeX}

\begin{document}

\begin{abstract}
\noindent
Stochastic resetting is prevalent in natural and man-made systems giving rise to a long series of non-equilibrium phenomena. Diffusion with stochastic resetting serves as a paradigmatic model to study these phenomena, but the lack of a well-controlled platform by which this process can be studied experimentally has been a major impediment to research in the field. Here, we report the experimental realization of colloidal particle diffusion and resetting via holographic optical tweezers. We provide the first experimental corroboration of central theoretical results, and go on to measure the energetic cost of resetting in steady-state and first-passage scenarios. In both cases, we show that this cost cannot be made arbitrarily small due to fundamental constraints on realistic resetting protocols. The methods developed herein open the door to future experimental study of resetting phenomena beyond diffusion.
\end{abstract}

\begin{tocentry}
\includegraphics{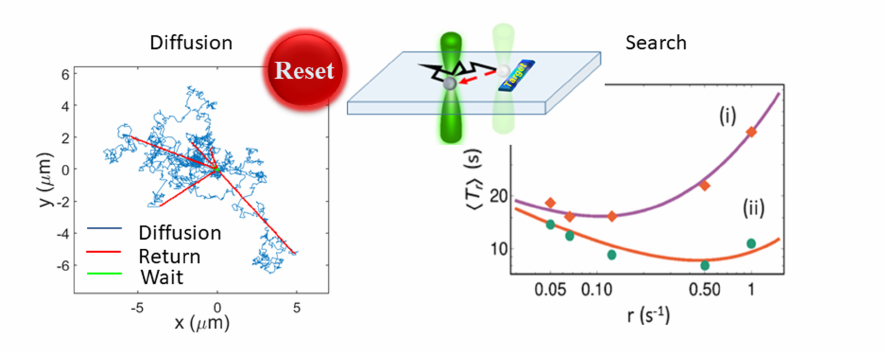}	
\end{tocentry}

\section{Introduction}

\noindent

Stochastic resetting is ubiquitous in nature, and has recently been the subject of vigorous studies \cite{review} in e.g., physics \cite{Evans2011_1,Evans2011_2,Evans2011_3}, chemistry \cite{Restart-Biophysics1,Restart-Biophysics2,Restart-Biophysics6}, biological physics \cite{Restart-Biophysics3,Restart-Biophysics8}, computer science \cite{restart-CS1,restart-CS2} and queuing theory \cite{queue1,queue2}. A stylized model to study resetting phenomena was proposed by Evans and Majumdar \cite{Evans2011_1}. The model, which considers a diffusing particle subject to stochastic resetting, exhibits many rich properties e.g., the emergence of a non-equilibrium steady state and interesting relaxation dynamics \cite{Evans2011_1,Evans2011_2,Evans2011_3,Evans2014_3,Pal2016_1,relaxation1,relaxation2,local} which were also observed in other systems with stochastic resetting \cite{restart_conc3,SEP,return1,return2,return3,return4,return5,Bod1,Bod2}. The model is also pertinent to the study of search and first-passage time (FPT) questions \cite{RednerBook,Schehr-review}. In particular, it was used to show that resetting can significantly reduce the mean FTP of a diffusing particle to a target by mitigating the deleterious effect of large FPT fluctuations that are intrinsic to diffusion in the absence of resetting \cite{Evans2011_1,Evans2011_2,Evans2011_3,review,Pal2016_1,Ray,interval,Durang2019}. Interestingly, this beneficial effect of resetting also extends beyond diffusion and applies to many other stochastic processes \cite{review,return5,Bod1,Bod2,return3,return4,ReuveniPRL,PalReuveniPRL,branching_II,Restart-Search1,Restart-Search2,Chechkin,Landau,HRS}; and further studies moreover revealed a genre of universality relations associated with optimally restarted processes as well as the existence of a globally optimal resetting strategy \cite{Restart-Biophysics1,Restart-Biophysics2,ReuveniPRL,PalReuveniPRL,Chechkin,branching_II,Landau,HRS}.
\begin{figure}[t!]
 \centering
\includegraphics[scale=0.6]{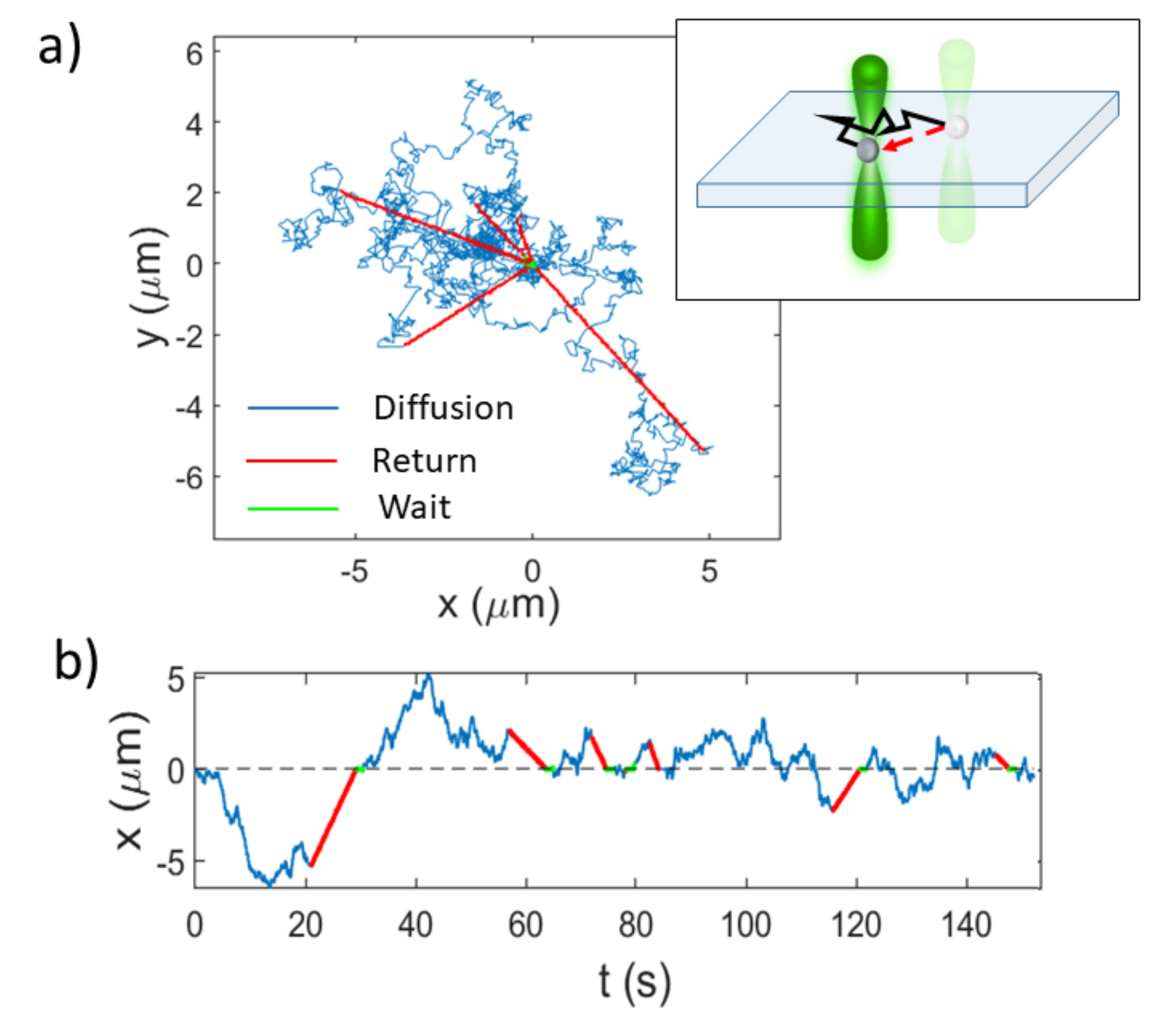}
\caption{Experimental realization of diffusion with stochastic resetting. a) A sample trajectory of a silica particle diffusing (blue) near the bottom of a sample cell. The particle sets off from the origin and is  reset at a rate $r=0.05 s^{-1}$. Following a resetting epoch, the particle is driven back to the origin at a constant radial velocity $v=0.8 \mu m/s$ using HOTs (red). After the particle arrives at the origin it remains trapped there for a short period of time to improve localization (green). Inset shows a schematic illustration of the experiment. b) Projection of the particle's trajectory onto the $x$-axis.}
\label{Fig:expt}
\end{figure}

Despite a long catalogue of theoretical studies on stochastic resetting, no attempt to experimentally study resetting in a controlled environment has been made to date, (but see very recent work that appeared after our arXiv submission \cite{Besga}). This is needed as resetting in the real world is never `clean' as in theoretical models which glance over physical complications for the sake of analytical tractability and elegance. In this letter, we report the experimental realization of diffusion with stochastic resetting (\fref{Fig:expt}). Our setup comprises of a colloidal particle suspended in fluid (in quasi-two dimensions) and resetting is implemented via  a home built holographic optical tweezers (HOTs) system \cite{Dufresne2001,Polin2005,Grier06,Crocker1996} described in  the Supporting Information accompanying this letter \cite{SM}. We study two, physically amenable, resetting protocols in which the particle is returned to the origin: (i) at a constant velocity, and (ii) within a constant time. In both cases, resetting is stochastic: time intervals between resetting events come from an exponential distribution with mean $1/r$.  


 Every experiment starts by drawing a series of random resetting times $\{t_1,t_2,t_3,...\}$ taken from an exponential distribution with mean $1/r$. At time zero, the particle is trapped at the origin and the experiment, which consists of a series of statistically identical steps, begins. At the $i$-th step of the experimental protocol, the particle is allowed to diffuse for a time $t_i$ eventually arriving at a position $(x_i,y_i)$. At this time, an optical trap is projected onto the particle and the particle is dragged by the trap to its initial position. A typical trajectory of a colloidal particle performing diffusion under stochastic resetting with $r=0.05s^{-1}$ is shown in \fref{Fig:expt}a and Supplementary movie 1 \cite{SM}. Note that the trajectory is composed of three phases of motion: diffusion, return, and a short waiting time to allow for optimal localization at the origin (\fref{Fig:expt}b). 

Below we utilize our setup to study the long time position distribution of a tagged particle and its dependence on the resetting protocol. We consider the energetic cost of resetting and characterize the mean and distribution of energy spent per resetting event.  Finally, we study the mean FPT of a tagged particle to a region in space and the energetic cost of resetting in this scenario. We conclude with discussion and outlook on the future of experimental studies of stochastic resetting.

\section{Stochastic resetting with instantaneous returns} 
\begin{figure}[t!]
\includegraphics[width=6.8cm,height=5.2cm]{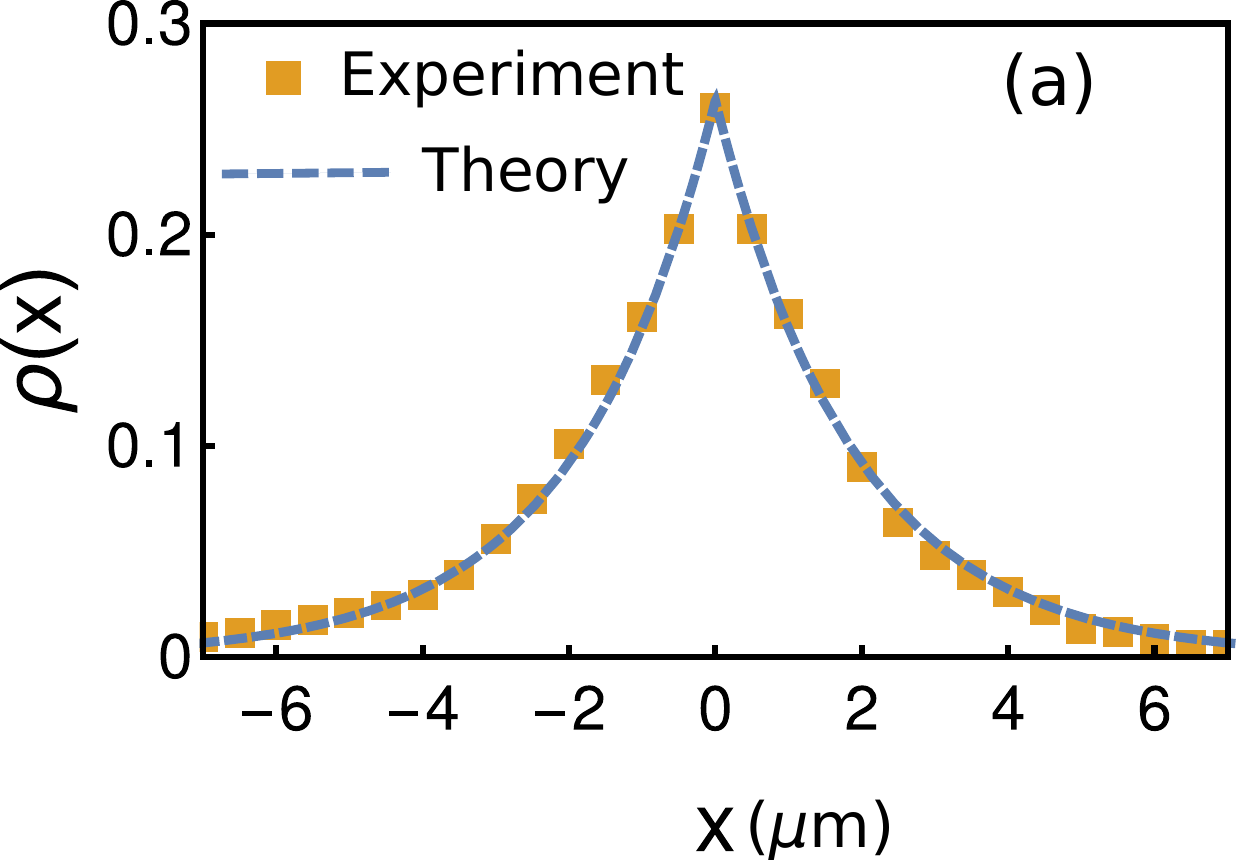}
\includegraphics[width=6.8cm,height=5.2cm]{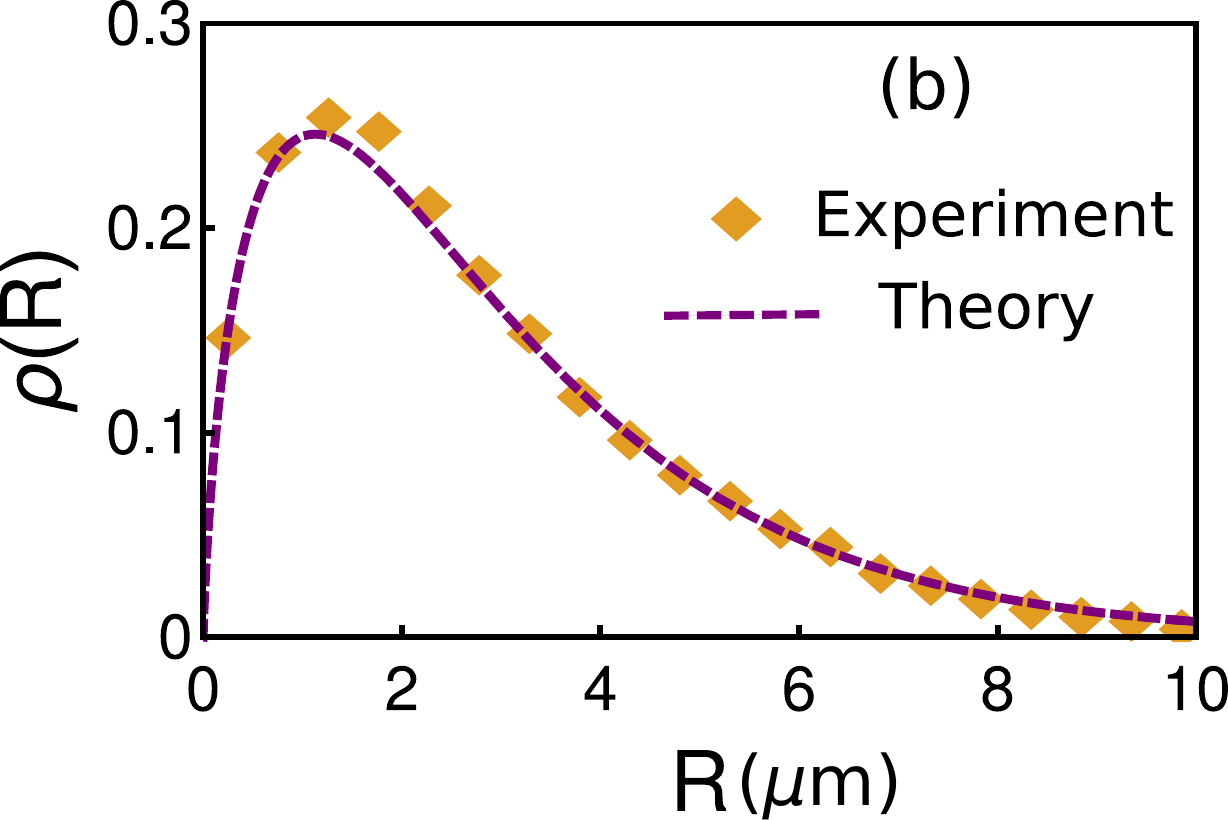}
	\caption{Steady-state distribution of diffusion with stochastic resetting and instantaneous returns. a) Distribution of the position along the $x$-axis. Markers come from experiments and the dashed line is the theoretical prediction $\rho(x)=\frac{\alpha_0}{2}e^{-\alpha_0|x|}$, where $\alpha_0=\sqrt{r/D}$, and $D$ is the diffusion constant. b) The radial position distribution. Markers come from experiments and the dashed line is the theoretical prediction   $\rho(R)=\alpha_0^2RK_0(\alpha_0R)$ \cite{SM} with $K_n(z)$ standing for the modified Bessel function of the second kind \cite{Stegun}. In both panels no fitting procedure was applied: $D=0.18\pm 0.02 \mu m^2/s$ was measured independently and $r=0.05s^{-1}$ was set by the operator.}  
	\label{Fig:ssDist}
\end{figure}

We first study the case in which upon resetting the particle is teleported back to the origin in zero time. This case was the first to be analysed theoretically \cite{Evans2011_1}, thus providing a benchmark for experimental results. 
A particle undergoing free Brownian motion is not bound in space. It has a Gaussian position distribution with a variance that grows linearly with time. Repeated resetting of the particle to its initial position will, however, result in effective confinement and in a non-Gaussian steady state distribution: $\rho(x)=\frac{\alpha_0}{2}e^{-\alpha_0|x|}$, where $\alpha_0=\sqrt{r/D}$, and $D$ is the diffusion constant\cite{Evans2011_1,Evans2011_2}.  Estimating the steady state distribution of the particle's position along the $x$-axis by digitally removing the return (red) and wait (green) phases of motion in \fref{Fig:expt}b \cite{SM}, we find that the experimentally measured results conform well with this theoretical prediction (\fref{Fig:ssDist}a). The steady state radial density of the particle can also be extracted from the experimental trajectories by looking at the steady-state distribution of the distance $R=\sqrt{x^2+y^2}$ from the origin. Here too, we find excellent agreement with theory (\fref{Fig:ssDist}b).

\section{Stochastic resetting with non-instantaneous returns}
\begin{figure}[t]
\includegraphics[width=13cm]{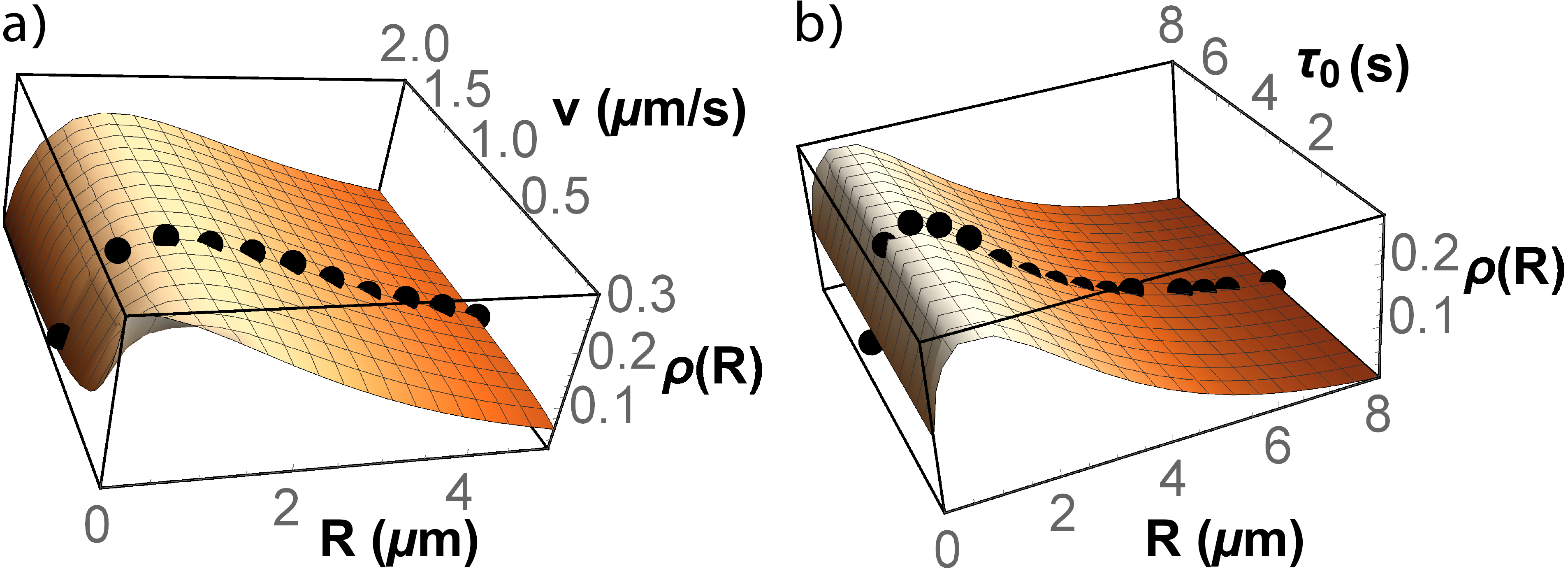}
\caption{Steady-state distributions of diffusion with stochastic resetting and non-instantaneous returns. a) The radial position distribution, $\rho(R)$, vs.  the distance $R$ and the radial return velocity $v$ as given by \eref{radial-constant-velocity}. Experimental results obtained for $v=0.8\mu m/s$ are superimposed on the theoretical prediction (black spheres). b) The radial position distribution vs. $R$ and the return time $\tau_0$ as given by \eref{radial-constant-time}. Experimental results obtained for  $\tau_0=3.79 s$ are superimposed on the theoretical prediction (black spheres).}
\label{non-inst}
\end{figure}

We now turn our attention to more realistic pictures of diffusion with stochastic resetting. These have just recently been considered theoretically in attempt to account for the non-instantaneous returns and waiting times that are seen in all physical  systems that include resetting \cite{Restart-Biophysics1,Restart-Biophysics2,Restart-Biophysics6,HRS,return1,return2,return3,return4,return5,Husain}. First, we consider a case where upon resetting HOTs are used to return the particle to the origin at a constant radial velocity $v=\sqrt{v_x^2+v_y^2}$ (\fref{Fig:expt}). This case naturally arises for resetting by constant force in the over-damped limit. We find that the radial steady state density is then given by \cite{SM}
\bea
\rho(R)=p_D^{c.v.}\rho_{\text{diff}}(R)+(1-p_D^{c.v.})\rho_{\text{ret}}(R),
\label{radial-constant-velocity}
\eea
where $p_D^{c.v.}=\left(1+\frac{\pi r}{2 \alpha_0 v} \right)^{-1}$ is the steady-state probability to find the particle in the diffusive phase. Here, $\rho_{\text{diff}}(R)=\alpha_0^2 R K_0(\alpha_0R)$ and $\rho_{\text{ret}}(R)=\frac{2\alpha_0^2}{\pi}RK_1(\alpha_0R)$ stand for the conditional probability densities of the particle's position when in the diffusive and return phases respectively, and $K_{n}(z)$ is the modified Bessel function of the second kind \cite{Stegun}. Bessel functions naturally appear here due to the rotational symmetry of the process and the resetting protocol. The result in \eref{radial-constant-velocity} is in very good agreement with experimental data as shown in \fref{non-inst}a and Fig. S3. We note that the theoretical result (\ref{radial-constant-velocity}) was also derived in \cite{2dreturn} using an alternative method.

Next, we consider a case where HOTs are used to return the particle to the origin at a constant time  $\tau_0$. This case is appealing due to its simplicity. Here, we find that the radial steady-state position distribution reads \cite{SM}
\bea
\rho(R)=p_D^{c.t.}\rho_{\text{diff}}(R)+(1-p_D^{c.t.})\rho_{\text{ret}}(R),
\label{radial-constant-time}
\eea
where $ p_D^{c.t.}=(1+r \tau_0)^{-1}$ is the steady-state probability to find the particle in the diffusive phase, and with  $\rho_{\text{diff}}(R)=\alpha_0^2 R K_0(\alpha_0R)$ and $\rho_{\text{ret}}(R)=\frac{\pi \alpha_0^2}{2}\left[ \frac{1}{\alpha _0}-R \left[K_0\left( \alpha _0 R \right) \pmb{L}_{-1}\left( \alpha _0 R \right) 
+K_1\left( \alpha _0 R \right) \pmb{L}_0\left( \alpha _0 R \right)\right] \right]$, standing for the conditional probability densities of the particle’s radial position when in the diffusive and return phases respectively. Here, $\pmb{L}_n$ is the modified Struve function of order $n$ \cite{Stegun}. The result in \eref{radial-constant-time} is in very good agreement with experimental data as shown in \fref{non-inst}b and Fig. S5.  
Note that \eref{radial-constant-velocity} and \eref{radial-constant-time} interpolate between the limit of instantaneous returns, with $v\rightarrow\infty$ or $\tau_0\rightarrow 0$, and the case of infinitely slow returns where $\rho(R)$ is dominated by the return statistics. Indeed, we find that  short return times and high return velocities are similar as returns are effectively instantaneous, while in the other extreme marked differences are observed (Fig. S4 and S6). 

\section{Energy cost per resetting}A central, and previously unexplored, aspect of stochastic resetting is the energetic cost associated with the resetting process itself. As discussed above, stochastic resetting prevents a diffusing particle from spreading over the entire available space as it normally would. Instead, a localized, non-equilibrium, steady-state is formed; but the latter can only be maintained by working on the system continuously.

\begin{figure}[t]
   \includegraphics[scale=0.6]{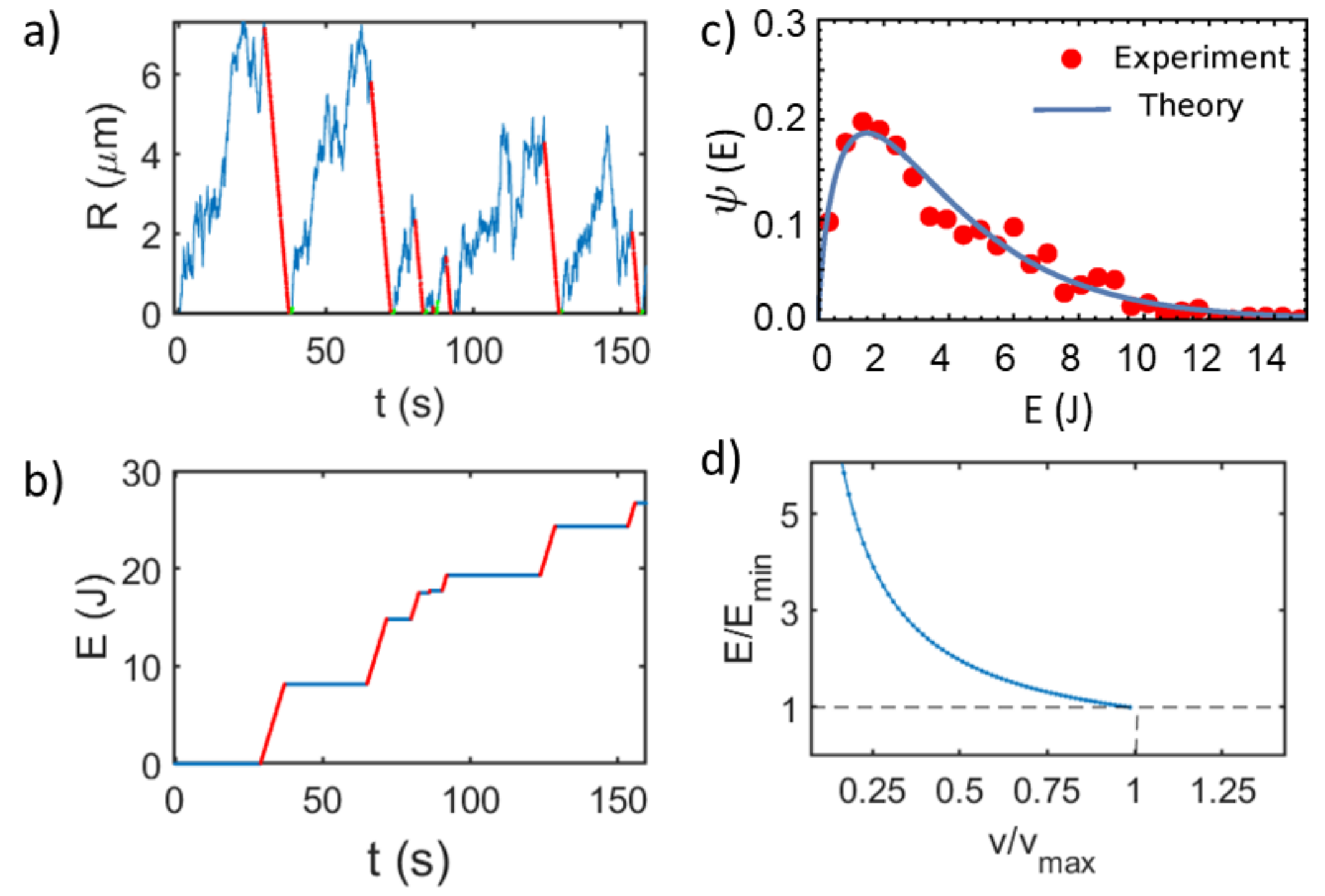}
\caption{Energetic cost of resetting. a) The radial distance from the origin vs. time for a particle diffusing with a  resetting rate $r=0.05s^{-1}$ and constant radial return velocity $v=0.8\mu m/s$.  b) The cumulative energy expenditure for the trajectory in panel a) (neglecting the cost of the wait period). c) The distribution of energy spent per resetting event. Red disks come from experiments and the theoretical prediction of \eref{Eq:Energy} is plotted as a solid blue line. d) Normalized energy spent per resetting event at constant power vs. the normalized radial return velocity as given by \eref{minmax}. The minimal energy is attained at a maximal velocity for which the trap is just barely strong enough to prevent the particle from escaping.}
\label{Fig:work}
\end{figure}

In our experiments, work is done by the laser to capture the particle in an optical trap and drag it back to the origin. The total energy spent per resetting event is then simply given by $E=\mathcal{P}\tau(R)$, where $\mathcal{P}$ is the laser power fixed at 1W and $\tau(R)$ is the time required for the laser to trap the particle at a distance $R$ and bring it back to the origin. As the particle's distance at the resetting epoch fluctuates randomly from one resetting event to another (\fref{Fig:work}a), the energy spent per resetting event is also random (\fref{Fig:work}b). To compute its distribution, we note that $E$ is proportional to the return time whose probability density function is in turn given by \cite{SM}
\bea
\phi(t)= \int_{\mathcal{D}}~d\vec{R}~\delta\left[t-\tau(\vec{R})  \right]\int_0^\infty dt'~f(t')~G_0(\vec{R},t').
\label{return_time1}
\eea
Here, $f(t)$ is the probability density governing the resetting time, $\vec{R}$ is the $d$-dimensional position vector, $\tau(\vec{R})$ is the return time and $G_0(\vec{R},t)$ is the propagator of the underlying stochastic dynamics. In our experimental setup, we have $f(t)=re^{-rt}$ and $G_0(R,t)=\frac{1}{4\pi Dt}e^{-R^2/4Dt}$ which is the diffusion propagator in polar coordinates. Moreover, in the case of constant radial return velocity $v$,  we have $\tau(\vec{R})=R/v$. A derivation then yields the probability density of the energy spent per resetting event \cite{SM} 
\bea
\psi(E)=\frac{E}{E_0^2}~K_0(E/E_0)~,
\label{Eq:Energy}
\eea
with $E_0=\alpha_0^{-1}v^{-1}\mathcal{P}$; and note that this is a special case of the K-distribution \cite{Redding,Long}. The mean energy spent per resetting event can be computed directly from \eref{Eq:Energy} and is given by $\langle E \rangle=\pi E_0/2$.  \eref{Eq:Energy} demonstrates good agreement with experimental data (\fref{Fig:work}c).

As $\langle E \rangle \propto v^{-1}$, it can be made smaller by working at higher return velocities. However, the stiffness, $k$, of the optical trap must be strong enough to oppose the drag force acting on the particle so as to keep it in the trap. Assuming the maximum allowed displacement of a particle in the trap is $\approx0.5\mu m$ \cite{Roichman07}, we find that working conditions must obey $k\ge2\gamma v$. As the stiffness is proportional to the laser power, $k=\mathcal{C} \mathcal{P}$ (where $\mathcal{C}$ is the conversion factor), the maximal working velocity is given by $v_{\text{max}}\approx \frac{1}{2} \mathcal{C} \mathcal{P}/\gamma$ which---independent of laser power---minimizes energy expenditure to  
$E_{\text{min}} \approx \pi \gamma \mathcal{C}^{-1} \alpha_0^{-1}$. Going to dimensionless variables we find  
\bea
\langle E \rangle/E_{\text{min}}=v_{\text{max}}/v, \hspace{0.5cm}
\label{minmax}
\eea
for $v<v_{\text{max}}$. This nicely illustrates that $\langle E \rangle$ cannot be lowered indefinitely, i.e., that there is a minimal energy cost per resetting event (\fref{Fig:work}d).


\section{Energy cost per first-passage}
\begin{figure}[t]
\includegraphics[scale = 0.6]{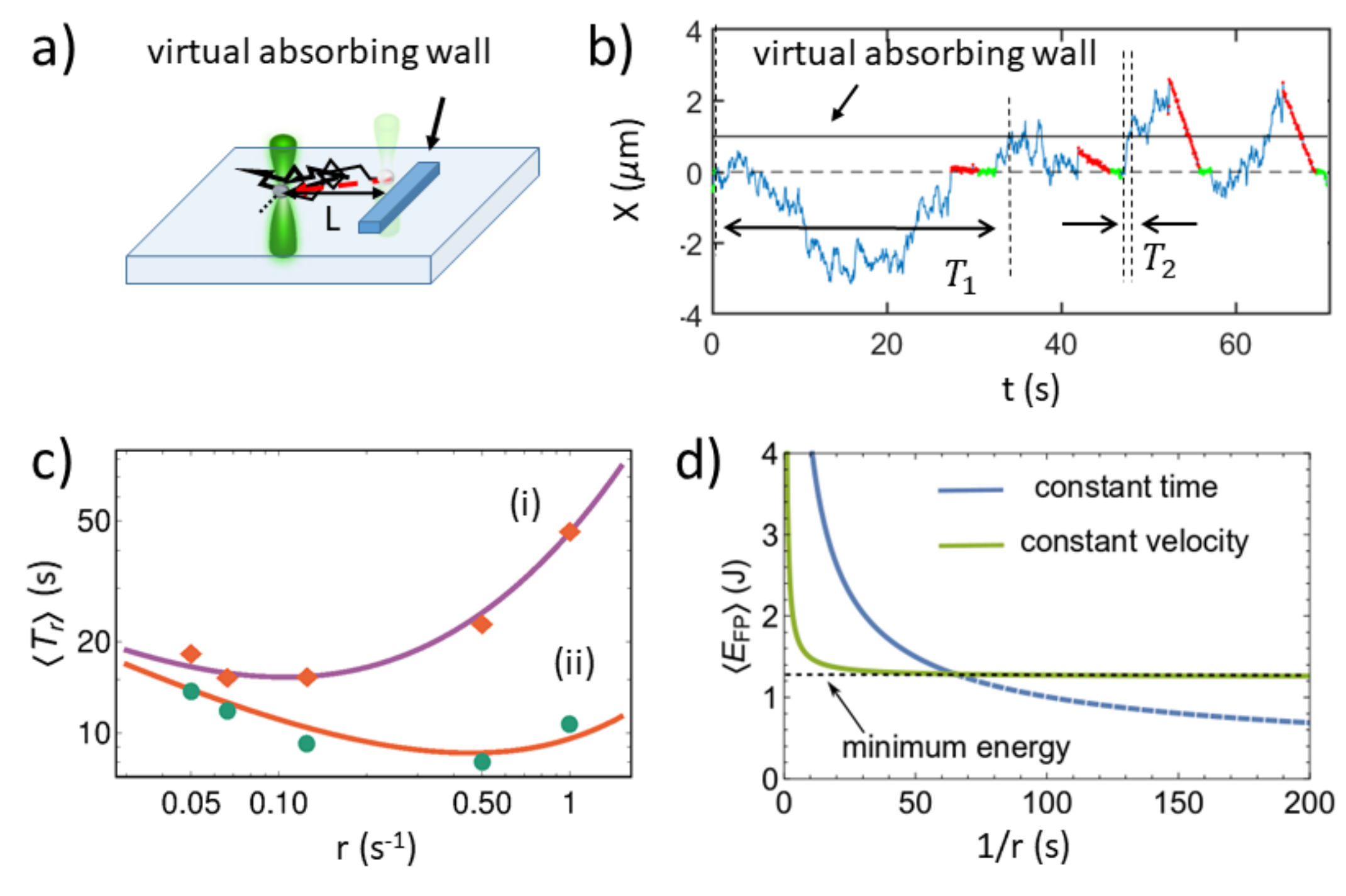} \caption{
a) Schematic illustration of a first-passage experiment. The target is a virtual absorbing wall. b) Projection of the particle's trajectory onto the $x$-axis. The position of the wall is marked as a solid line. Returns are marked in red (return time $\tau_0=3.79s$) and waiting periods are marked in green. The first two first-passage times are marked by $T_1$ and $T_2$. c) The mean FPT to the wall vs. the resetting rate for: (i) non-instantaneous returns with $\tau_0=3.79s$; and (ii) instantaneous returns ($\tau_0=0$). Theoretical predictions [\eref{Eq:MFPT_return}, solid lines] are in good agreement with experimental data (symbols). d) Energy cost per first-passage event for the constant time (blue, $\tau_0=3.79s$) and constant velocity (green, $v=0.8\mu m/s$) return protocols. For a fixed laser power, the energetic cost of the constant velocity protocol is bounded from below. In the constant time protocol, we have dashed the range of resetting rates where the average return velocity is higher than $v$ \cite{SM}. The latter cannot exceed $v_{\text{max}}$, thus bounding the energy cost of the constant time protocol.}
\label{Fig:MFPT}
\end{figure}

Having looked at stationary properties of diffusion with resetting, we now turn attention to first-passage properties which have numerous applications \cite{Evans2011_3,Restart-Biophysics1,Restart-Biophysics2,Restart-Biophysics6,Restart-Biophysics3,Restart-Biophysics8,RednerBook,Schehr-review,ReuveniPRL,PalReuveniPRL,branching_II,Restart-Search1,Restart-Search2,Chechkin,Landau,HRS,Frinkes2010,Branton2010,Tu2013,Bezrukov2000,Grunwald2010,Ghale2014,Ma2013,Iyer2014,Ingraham1983,Amir2014,Osella2014,cooper1991,MetzlerBook}. We recall that while the mean first-passage time (MFPT) of a Brownian particle to a stationary target diverges \cite{RednerBook,Schehr-review}, resetting will render it finite \cite{Evans2011_1}, even if returns are non-instantaneous \cite{Restart-Biophysics1,Restart-Biophysics2,Restart-Biophysics6,HRS,return3,return5}. To experimentally show this, we consider the setup in \fref{Fig:MFPT}a.

A first passage experiment starts at time zero when the particle is at the origin. Resetting is conducted stochastically with rate $r$, and HOTs are used to return the particle to the origin at a constant return time $\tau_0$. However, we now also define a target, set to be a virtual infinite absorbing wall located at $x=L$, i.e., parallel to the $y$-axis. The particle is allowed to diffuse with stochastic resetting until it hits the target, and the hitting times (FPTs) are recorded (Fig.~\ref{Fig:MFPT}b). A typical trajectory extracted from such an experiment with $\tau_0= 3.79s$, $L = 1 \mu$m, and $r=0.05s^{-1}$ is shown in Fig.~\ref{Fig:MFPT}b, Fig.~S7, and Supplementary movie 2. Measurements were also taken for: $r=0.0667s^{-1}$, $0.125s^{-1}$, $0.5s^{-1}$, and $1s^{-1}$ \cite{SM}.

To check agreement between experimental FPT data and theory, we derived a formula for the mean FPT of diffusion with stochastic resetting and constant time returns \cite{SM}
\bea
\langle T_r \rangle=\left( \frac{1}{r}+\tau_0 \right) \left[ e^{\sqrt{rL^2/D}}-1  \right].
\label{Eq:MFPT_return}
\eea
\eref{Eq:MFPT_return} is in excellent agreement with data as shown in Fig.~\ref{Fig:MFPT}c, including accurate prediction of the optimal resetting rate which minimizes the mean FPT of the particle to the target. 

As resetting requires energy, lowering the mean FPT will have a cost---which to date has been completely ignored. 
To compute it, we require the probability density of the return time in a FPT scenario which is generally given by \cite{SM}
\bea
\phi_{\text{FP}}(t)=
\frac{1}{p}\int_{\mathcal{D}}~d\vec{R}~\delta[t-\tau(\vec{R})]~\int_0^\infty~dt'~f(t')~G_{\text{abs}}(\vec{R},t'),~~~
\label{return_time2}
\eea 
where $p$ is the probability that a reset event will occur before a first passage event 
and $G_{\text{abs}}(\vec{R},t)$ is the reset-free propagator in the presence of the absorbing target. As the number of resets per first passage event is geometrically distributed with mean $p/(1-p)$, one can compute, $\langle  E_{\text{FP}} \rangle$, the average energy spent per first passage event \cite{SM}. Setting $\tau(\vec{R})=\tau_0$, we find 
$\langle  E_{\text{FP}} \rangle= \mathcal{P}\tau_0 \left(e^{\sqrt{rL^2/D}}-1\right)$  which vanishes as $r \to 0$ (\fref{Fig:MFPT}d) \cite{SM}. Note, however, that in this limit $|\vec{R}|$ can be very large at the resetting moment which inevitably implies frequent cases where $|\vec{R}|/\tau_0>v_{\text{max}}$. This in turn results in particles escaping the optical trap and in utter breakdown of the constant return time protocol \cite{SM}. To avoid this problem, we instead consider the more realistic  constant velocity protocol which gives $\langle  E_{\text{FP}} \rangle= \frac{ \mathcal{P} L }{v} ~\left[\frac{2\sinh{\alpha_0 L}}{\alpha_0 L}-1\right]$, for $v<v_{\text{max}}$ (\fref{Fig:MFPT}d). This result surprisingly reveals a dynamical transition: while $\langle E_{\text{FP}} \rangle \equiv 0$ when $r=0$, for all $r>0$ one has $\langle E_{\text{FP}} \rangle > \mathcal{P} L/v $, which means that the energy spent per FPT event cannot drop below that which is required to drag the particle directly to $L$ at a constant velocity $v$. Setting $v=v_{\text{max}}$ in the above bound gives $\langle E_{\text{FP}} \rangle > 2L\gamma/ \mathcal{C}$, which does not depend on laser power or return velocity.

\section{Discussion and future outlook}In this study, we have demonstrated a unique and versatile method to realize experimentally a resetting process in which many parameters can be easily controlled.  To test our platform, we first used it to experimentally corroborate existing theoretical predictions, which in turn motivated experimental and theoretical study of novel and more realistic aspects of diffusion with stochastic resetting. Of prime importance in this regard is the energetic cost of resetting \cite{thermo1,thermo2,thermo3}, which we have characterized in both the steady-state and first-passage settings. Combining analytically derived expressions with the physics of resetting via HOTs then surprisingly revealed lower bounds on the energy spent per resetting for steady state and first passage events. Our results were based on Eqs. (\ref{return_time1}) and (\ref{return_time2}) which are general and can be used as platform to extend our findings to a wide range of stochastic motions, resetting time distributions, return protocols, and arbitrary dimensions. In addition, our setup can be easily adapted to experimentally explore regimes that are well beyond the reach of existing theories of stochastic resetting, e.g.,  multibody systems with strong interactions. These will be considered elsewhere. 

Finally, we note that the optical trapping method used herein is far from being the most efficient way to apply force to a colloidal particle. In fact, in our experiments we used $1W$ of power at the laser output to create a trap of $k=30pN/\mu m$ for a silica bead of radius $a=0.75\mu m$. For experiments with a constant return velocity $v=0.8\mu m/s$ and resetting rate $r=0.05s^{-1}$, the average return time was $ \langle \tau(R) \rangle=\pi \alpha_0^{-1}v^{-1}/2=3.68s$, where the average was done with respect to $\phi(t)$ using \eref{return_time1}. This translates to an average energy expenditure of $\langle E\rangle=\mathcal{P}\langle \tau(R) \rangle=3.68\pm0.05J$ per resetting event. In contrast, the work done against friction to drag the particle at a constant velocity $v$ for a distance $R$ is given by $W_{\text{drag}}=\gamma v R$ where  $\gamma=6\pi\eta a$ is the the Stokes drag coefficient. Taking averages, we find $\langle R \rangle=v\langle \tau(R) \rangle=\pi \alpha_0^{-1}/2 $. The work required per resetting event is then given by $\langle W_{\text{drag}} \rangle= \gamma v \langle R  \rangle = \pi \alpha_0^{-1} \gamma v/2$, which translates into $3.4\cdot10^{-20}J$ or $8.3 k_BT$ per resetting event. We thus see that $\langle W_{\text{drag}} \rangle \ll \langle E \rangle$, i.e., that the work required to reset the particle's position is orders of magnitude smaller than the actual amount of energy spent when resetting is done using HOTs. Developing energy efficient resetting methods is a future challenge.

\begin{acknowledgement}

The authors acknowledge Gilad Pollack for his help in coding the resetting protocol of the HOTs. A. P. acknowledges support from the Raymond and Beverly Sackler Post-Doctoral Scholarship at Tel-Aviv University; and Somrita Ray for many fruitful discussions. A. S. acknowledges support from the Ratner center for single molecule studies.  S. R. acknowledges support from the Azrieli Foundation, from the Raymond and Beverly Sackler Center for Computational Molecular and Materials Science at Tel Aviv University, and from the Israel Science Foundation (grant No. 394/19). Y. R. acknowledges support from the Israel Science Foundation (grant No. 988/17).

\end{acknowledgement}

\begin{suppinfo}

See Supplementary Information for the details of the theoretical derivations, experimental methods, and other results.

\end{suppinfo}

\section{Author contributions}
Ofir Tal Friedman, Arnab Pal and Amandeep Sekhon have equally contributed to this work.


\end{document}